\documentclass[conference,10pt]{IEEEtran}
\usepackage{caption}
\usepackage{subcaption}

\usepackage{booktabs} 
\usepackage{graphicx}
\usepackage{multirow}
\usepackage{marginnote,soul}
\usepackage{todonotes}
\usepackage[normalem]{ulem}
\useunder{\uline}{\ul}{}

\usepackage[ruled,vlined]{algorithm2e}
\SetAlFnt{\footnotesize}
\SetAlCapFnt{\footnotesize}
\SetAlCapNameFnt{\footnotesize}

\usepackage{lipsum}
\usepackage{amssymb}
\usepackage{bbm}

\usepackage{enumitem}

\usepackage{amssymb,amsmath}
\usepackage{amsmath}

\usepackage{color}
\usepackage{booktabs}

\usepackage{tikz}
\usetikzlibrary{shapes.geometric, arrows}
\usetikzlibrary{fit}
\tikzstyle{startstop} = [rectangle, rounded corners, minimum width=3cm, minimum height=1cm,text centered, draw=black, fill=red!20]
\tikzstyle{io} = [trapezium, trapezium left angle=70, trapezium right angle=110, minimum width=3cm,  minimum height=1cm, text centered,text width=3cm, draw=black, fill=blue!20]
\tikzstyle{process} = [rectangle, minimum width=3cm, minimum height=1cm, text centered, text width=3cm, draw=black, fill=orange!20]
\tikzstyle{decision} = [diamond, minimum width=3cm, minimum height=1cm, text centered, draw=black, fill=green!20]
\tikzstyle{arrow} = [thick,->,>=stealth]

\title{Recommendation System-based Upper Confidence Bound for Online Advertising}
\author{%

\IEEEauthorblockN{Nhan Nguyen-Thanh$^{1,2}$, Dana Marinca$^{2,1}$, Kinda Khawam$^{2,1}$,\\
David Rohde$^{3}$,  Flavian Vasile$^{3}$, 
Elena Simona Lohan$^{4}$, 
 Steven Martin$^1$, Dominique Quadri$^1$}

\IEEEauthorblockA{
$^1$ LRI, University Paris-Saclay, Paris-Sud, France\\ 
$^2$ David Lab., University Paris-Saclay, UVSQ, France\\
$^3$ Criteo, France\\
$^4$ {Tampere University, Finland}\\
Email: nhan.nguyen@lri.fr,dana.marinca@uvsq.fr,   
kinda.khawam@uvsq.fr,  \\
d.rohde@criteo.com, f.vasile@criteo.com,
elena-simona.lohan@tuni.fi,  \\
steven.martin@lri.fr , 
dominique.quadri@lri.fr} 
}

\begin{document}

\maketitle
\begin{abstract}
In this paper, the method UCB-RS, which resorts to recommendation system (RS) for enhancing the upper-confidence bound algorithm UCB, is presented. The proposed method is used for dealing  with non-stationary and large-state spaces multi-armed bandit problems. The proposed method has been targeted to the problem of the product recommendation in the online advertising. Through extensive testing with RecoGym, an OpenAI Gym-based reinforcement learning environment for the product recommendation in online advertising, the proposed method outperforms the widespread reinforcement learning schemes such as $\epsilon$-Greedy, Upper Confidence (UCB1) and Exponential Weights for Exploration and Exploitation (EXP3).
\end{abstract}

\begin{IEEEkeywords}
Recommendation System, multi-armed bandit, OpenAI Gym, RecoGym,  Epsilon Greedy, EXP3, EXP3S, UCB, UCB1, UCB-RS
\end{IEEEkeywords}

\IEEEpeerreviewmaketitle
\section{Introduction} \label{sec:Intro}

Online advertising is becoming increasingly popular and is the main motivation for the development of almost free internet platforms such as search engines, social networks, recruitment sites, multimedia contents (e.g., videos, images, musics, ...) sharing, etc. From the point of view of the internet users, the product recommendation on online advertising can be genuinely useful if it meets the real immediate needs of users. Instead of spending a lot of time and effort searching for a huge number of thousands or even millions of choices, most internet users will be
quite satisfied if recommendation systems propose exactly what they need. Finding a good recommendation system, therefore, continues to be the goal of many studies \cite{Rohde2018,Bresler2018}.

Online and offline approaches for learning optimal recommendation policies can be found in the literature. On one hand, a product recommender in the offline approach is basically a \textit{recommendation system} (RS) which could be a classical classification/regression algorithm or a modern matrix completion method. The recommendation systems in this approach focus mainly on exploiting historical data for the next item prediction, and the main shortcoming of this approach is the mismatch between the offline metrics and the real online performances \cite{Rohde2018}. 

On the other hand, the product recommendation in the online learning approach is considered as a \textit{reinforcement learning} (RL)/ \textit{multi-armed bandit} (MAB) problem \cite{Kohli:2013:FBA:2891460.2891618,Mary:2015:BRS:2942197.2942226}. 
Online algorithms solving RL/MAB problems, such as $\epsilon$-greedy \cite{Tokic:2010:A9E:1882150.1882177}, UCB \cite{Auer2002}, EXP3 \cite{Auer2003}, 
etc., select items and make suggestions to a user based on his/her current context (e.g., time of day, place, historical activities, etc.). They  must optimize the balance between exploiting known policies and exploring randomized ones for obtaining good performances. The drawbacks of the online approach are the lack of suitable datasets which contain bandit feedback and the high cost of implementing real tests. 

When adopting online learning to product recommendation, one has to deal with two challenges: the non-stationary characteristics and the large size of the space of the product recommendation environment. Indeed, the target of the online learning algorithm is a human being user whose behaviour is almost unpredictable, and the action space of online learning algorithms could be up to millions of items. This makes conventional RL algorithms almost inapplicable because the natural length of an advertising is very small compared to the action space. In the literature, few efforts on adopting online learning to product recommendation have been made. For instance, some modifications of the RL algorithm is found in \cite{Garivier2008}, by using discounted and sliding-window memory for coping with the non-stationary problem, whereas the work in \cite{Bresler2018} consider RS in a sequential process as an RL algorithm. However, due to the lack of suitable environment or dataset, the work in this direction still suffer from many limitations on optimizing parameters automatically. 

RecoGym \cite{Rohde2018}, a reinforcement learning environment for the problem of product recommendation in online advertising developed based on OpenAI Gym \cite{openAIGym}, introduces a combination of the two natural types of user-item interactions, which are the interaction of the user with the items on e-commerce websites and the interaction of the user with the ads of items on publisher websites, in only one simulator. This provides us a solution to overcome the above cited issues of RS and RL, and enable us to jointly consider both the recommendation system approach and the reinforcement learning approach for product recommendation at the same time.

In this paper, we utilize RecoGym as the online advertising environment for investigating our solution. We enhance the upper-confidence bound algorithm by utilizing user-user collaborative filtering for dealing with  non-stationary and large action spaces challenges of the product recommendation problems. By utilizing a reference set which contains historical data of several users, we estimate the possible achieved reward from a targeted user for each product based on his/her online interactions. The feedback of the targeted user on the recommended product, which has the highest confident bound computed based on the estimated reward and a confidence level, will be repeatedly used for the above estimation. In this way, the RS and UCB will be combined, and deemed \textit{UCB-RS} algorithm.  

The rest of the paper is presented as follows. In section \ref{sec:Problem}, we introduce the problem and explain in detail the framework of RecoGym. In section \ref{sec:MultiArmedAlgs}, we recall some baseline MAB algorithms. In section \ref{sec:Propose}, we describe our solution. In section \ref{sec:Result}, we present the numerical results. Finally, in section \ref{sec:Conclusion}, we conclude the paper.

\section{Problem description} \label{sec:Problem}
\subsection{Product recommendation problem}

\begin{figure}[htbp]
\includegraphics[trim=0.5cm 9.5cm 0.5cm 2.3cm,clip, width = \linewidth]{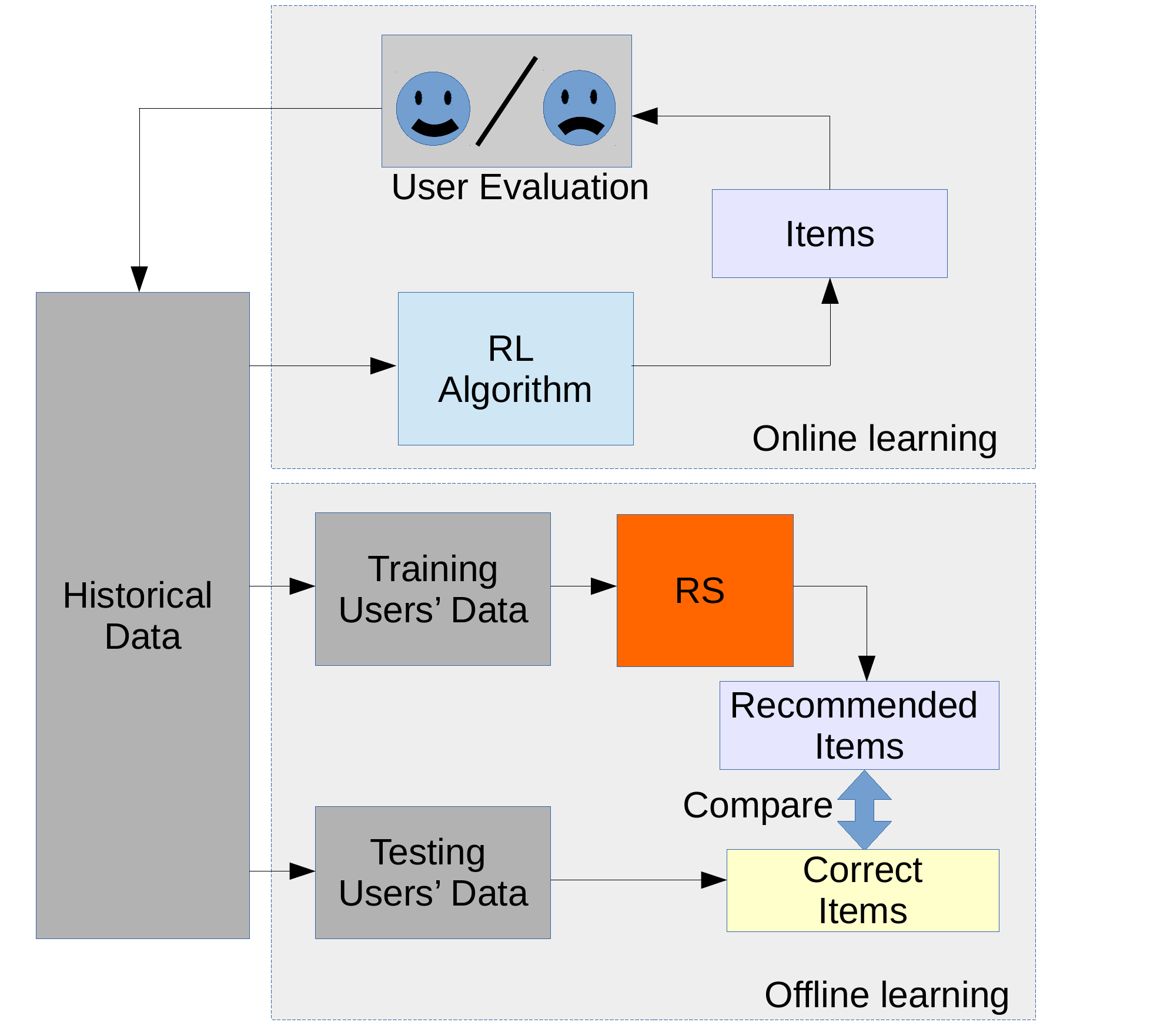}
\caption{Online and offline learning approaches in product recommendation problems}
\label{fig:OnlineOffline}
\end{figure}

Generally, there are two approaches for learning optimal recommendation policies: the conventional offline learning and the online learning as shown in Fig. \ref{fig:OnlineOffline}.  
In the offline approach, a product recommender could be a classical classification/regression algorithm or a modern matrix completion method that exploit historical data for the next item prediction. The associated offline metrics could be mean squared error, precision, area under the curve, etc. However, there are more and more reports regarding the mismatch between offline metrics and online real performance.

In the online learning approach, the product recommendation is considered as a reinforcement learning/multi-armed bandit problem where the online algorithm will select an item to make relative suggestions to the user based on his/her current context such as time of day, place, historical activities, etc. Online learning algorithms, such as $\epsilon$-greedy or UCB,  must optimize the balance between exploiting known policies and exploring randomized ones for obtaining good performances. The main drawback of such approaches is the high cost of implementing online testing and the lack of suitable datasets that includes bandit feedback to arbitrary recommendation policies.

For online advertising, there are two types of interaction that the user can have with items:(i) the browsing of user on items displayed on e-commerce websites, called \textit{organic sessions} and (ii) the interaction of user on items displayed on the ads of the publisher websites, called \textit{bandit sessions}\cite{Rohde2018}. Obviously, the data acquired on organic sessions and the feedback received on bandit sessions allow us to combine both online and offline approaches. The latter combination is exploited by \textit{RecoGym}, a reinforcement learning environment presented in detail in the next section, aiming for a better product recommendation system.


\subsection{RecoGym} 
RecoGym \cite{Rohde2018} is a reinforcement learning environment for the problem of product recommendation in online advertising developed based on OpenAI Gym environment \cite{openAIGym}. The simulator includes both the organic and the bandit user-item interactions. It allows us to adjust (i) the correlation between the two types of interactions, (ii) the dimension of the hidden space of user interests and items characteristics, and (iii) the level of relative impact of the user’s past exposure level to ads on the click-through rate of an ad display at a given time.
Similar to any OpenAI Gym environment, RecoGym procceds in the following manner:
\begin{itemize}
    \item \textit{\textbf{Reset}} - The simulator creates a random synthetic user when it is called. 
    \item \textit{\textbf{Step}} - At each step call, the simulator receives the input action, i.e., the recommended item, from the agent and return four objects:
    \begin{itemize}
        \item \textit{Observation}: returns (None) if the user session is of bandit type, and returns all the viewed items if the user session is of organic type.
        \item \textit{Reward}: returns (click/non-click) of the environment for the input item.
        \item \textit{Done}: returns (True) if the user has finished his/her shopping sequence and vice versa. If true, the Reset function will be called for initializing a new user for the next step.
        \item \textit{Info}: contains possible logging information from the simulation.
    \end{itemize}
\end{itemize}
After executing a \textbf{reset}, the \textbf{step} call generates the first organic batch of the new user. The action and the reward only appear in the subsequent calls of steps.

\section{Multi-armed bandit algorithms} 
\label{sec:MultiArmedAlgs}
In this section, we briefly present the baseline algorithms for handling the MAB (Multi-Armed Bandit) problem which is considered in the paper. They include $\epsilon$-greedy, UCB1 (Upper Confidence) and EXP3 (Exponential Weights for Exploration and Exploitation).
\subsection{$\epsilon$-Greedy algorithm}
The $\epsilon$-greedy algorithm is widely used to solve the multi-armed bandit problem owing to its simplicity. The only parameter of the method is $\epsilon$. At each round $t$, we select the best arm with a probability $1-\epsilon$ or we select a random arm with probability $\epsilon$. The value of $\epsilon$ is fixed in the standard settings. The $\epsilon$ value can be changed with time. Accordingly, $\epsilon$ decreases over time to ensure that the algorithm explores more at the initial rounds and less in further ones.

\subsection{Upper-Confidence Bound algorithm \cite{Auer2002}}
In $\epsilon$-greedy algorithm, the operation focuses on exploiting the best arms and exploring random arms. There is no tracking about the knowledge reliability relative to arms. Instead, UCB pays attention to not only the arms values but also to arms confidence. 

Keeping track of confidence in the values of arms is important because the returned rewards from selected arms are basically noisy. Using historical data of an arm to estimate its values always introduces noise. A larger size of historical data can help reduce the estimation noise and hence improves confidence interval.

Upper Confidence Bounds (UCB) algorithms use the naive mean estimator for arms reward and a confidence interval defined based on the size of the past experience of arms. Such algorithms are considered as an exploration adjustment to the empirical mean. At each step, UCB algorithm follows the “optimism-in-face-of-uncertainty” principle to select the arm with the highest upper confidence bound, defined by \eqref{Eqn:UCB}.

\begin{equation}
U_{k,t} = \mu_i + \sqrt{\frac{\alpha \log t}{N_i}}
\label{Eqn:UCB}
\end{equation}
where $\mu_i$ is the mean of arm $i$, $N_i$ is the number of times that arm $i$ is played up-to time $t$, and $\alpha > 0$. It can be seen that Upper-Confidence Bound of an arm is composed by the estimated mean reward and the confidence level of the arm. The UCB algorithm is sketched in Alg. \ref{alg:UCB1}. It can be seen that there are two stages. In the first stage, UCB algorithm plays each arms once. This stage corresponds to a purely exploration phase which ensures some initial value for calculating confidence bound for all arms. In the second stage, the upper confidence bound value of each arm will be computed. Simple mean estimator is used to calculate the estimated reward of arms.

\begin{algorithm}[t]
\SetAlgoLined
\For{$t\leftarrow 1$ \KwTo $K$}{
\emph{Play arm $I_t = t$}\;}
\For{$t\leftarrow N+1$ \KwTo $T$}{
\emph{Play arm $I_t = \underset{k}\arg\max (U_{k,t})$}\;}

\caption{UCB1}
\label{alg:UCB1}
\end{algorithm}

\subsection{EXP3 and EXP3S algorithms}
An approach befitting the non-stationary MAB problem is a variant of the \textit{ex}ponential-weight algorithm for \textit{ex}ploration and \textit{ex}ploitation (EXP3) algorithm \cite{Auer2003}. EXP3 was developed for adversarial bandits. It uses an egalitarian factor $\gamma \in [0,1]$ for adjusting the fraction of time the algorithm picks a purely random action. The rest of the time, the algorithm utilizes a list of weights for each actions to decide the probabilities of arms based on which the next action will be selected. The final probability of playing arm $i$ at time $t$ is given by:
\begin{equation}
p_{i,t} = (1-\gamma)\frac{w_{i,t}}{\sum_{j=1}^K w_{j,t}} +  \frac{\gamma}{K} 
\label{Eqn:pEXP3}
\end{equation}
where $K$ is number of arms, and $w_{i,t}$ is the weight of arm $i$ at time $t$.
At each time $t$, the weight of the selected arm is updated based on an exponential rule given by \eqref{Eqn:weightEXP3} for EXP3 and \eqref{Eqn:weightEXP3S} for EXP3S
\begin{subequations}
\begin{align}
\label{Eqn:weightEXP3}
w_{i,t+1} &= w_{i,t}\exp{\left(\gamma \frac{\rho_{i,t}}{p_{i,t}K} \right)} \\
\label{Eqn:weightEXP3S}
w_{i,t+1} &= w_{i,t}\exp{\left(\gamma \frac{\rho_{i,t}}{p_{i,t}K} \right)} 
 + \frac{e\delta}{K}{\sum_{j=1}^K w_{j,t}}
\end{align}
\end{subequations}
where $\delta>0$ and $\rho_{i,t}$ is the reward of arm $i$ at time $t$.

\section{The proposed UCB-RS} \label{sec:Propose}
In this section, we present our proposed Recommendation system-based UCB (UCB-RS) which is a modification version of UCB1 for dealing with both challenging issues: \textit{non-stationary} and \textit{large action spaces} of MAB problems.

It has been shown that UCB1 policies are rate optimal in the stationary case \cite{Garivier2008}, where the rewards distribution do not vary in time. However, in non-stationary cases, the bandit algorithm has to deal with a varying environment. The simple mean estimator of UCB1 can not be a good approach for the estimated reward. Hence, modifying the UCB estimator is necessary for such a case. The discounted UCB \cite{Hartland2006} and the sliding-window UCB \cite{Garivier2008} are empirical examples of the UCB modifications for coping with the first challenging non-stationary environment. 

For the second challenge, the bandit algorithm has to deal with a large arms space (e.g. thousand or even million arms-. In this case, the purely exploring stage becomes a heavily time-consuming process. UCB algorithm is almost inapplicable, especially for short playing time, i.e., $T<<K$, bandit environment.

Advertisement problems, where products are recommended for users to click and view, comprise unfortunately both of the above issues. Indeed, the environment of MAB is the user who has his/her own latent variables of interest on products, and these variables change over time and user's characteristics. Furthermore, depending on the user's habits, the length of playing an advertising session could be very short compared to the total number of products needed to be advertised. 
As a result, fitting UCB to such a case is vital. Because the modifications mainly target the starting phase of UCB, we call it the \textit{warming-start UCB}. 

The idea for our proposed modification for UCB is to use a recommendation system for estimating the arms' rewards. Utilizing the complementing properties of the recommendation system, the proposed algorithm takes into account the information of correlating arms with a very sparse input.
For example, with only a click on a laptop as the input, recommendation system can easily infer that the user will be interested in the products on the same category like headphones, mouses, speakers, etc., by considering the historical interaction of others user. Hence, it can estimate higher possible rewards for those products. By this way, the estimated arms rewards could be improved.
\subsection{Recommendation System Framework}  
\label{sec:RSFramework}
Recommendation system (RS) enables us to predict the most appropriate items by handling a large amount of data \cite{4654410}. A wide range of applications such as e-commerce frameworks, e.g.,  Amazon, Netflix \cite{leskovec2014mining}, social networks, e.g., Google New Personalization System, Facebook Friend recommendation \cite{aggarwal2016recommender}, etc. has been solved by using RS. Two main approaches taken in RS are content filtering and collaborative filtering. Content filtering utilizes auxiliary information associated with items and users such as location, age, gender, etc. Inversely, collaborative filtering (CF) works on observed user preferences. Therefore, despite their different profiles, two users are believed to be similar if their preferences are similar. Also, two items are considered as similar if the preferences given for them by most of users are similar. In general, collaborative filtering is equivalent to performing matrix completion for the sparse preference matrix. The problem of matrix completion is to estimate the rest of the matrix so that it satisfies some conditions with a subset of the observed entries.
In this way, RS becomes very effective in dealing with the product recommendation problem given a very sparse input information. 

It should be noticed that when deploying a RS to an environment with online feedback, each proposal affects what is learned about the users and items, thereby determining the possible accuracy of future proposals. Therefore, making a good RS for such the environment requires the same optimizing trade-off between exploration and exploitation of MAB problems. This empowers the idea of combining RS and MAB algorithm with a high potential.

Since the MAB algorithm runs on a specific user, CF, which considers at the same time multiple users-items relations, is tailored for the combination. The goal of CF in this case is to obtain as much as possible information about the target user preferences. Therefore, we use the user-user CF framework which is described as follows.

There are many metrics for measuring the similarity such as Euclidean, Cosine, Peason, Jaccard, Manhattan distances. In this paper, we select cosine distance, which provides good performance for sparse inputs, for measuring the similarity between users.
%
%
Given features vectors $\textbf{u}$ and $\textbf{v}$ of user $U$ and user $V$, the cosine similarity between user $U$ and user $V$ is defined by:
\begin{equation}
    sim(\mathbf{u},\mathbf{v})=\frac{\left \langle \boldsymbol{\mathbf{u}},\boldsymbol{\mathbf{v}} \right \rangle}{\left \| \left | \boldsymbol{\mathbf{u}} \right | \right \|\left \| \boldsymbol{\mathbf{v}} \right \|}
    \label{eqn:cosineSim} 
\end{equation}

After computing cosine similarity between user $U$ and all of the users in the reference set $R$, we select a subset $\mathcal{R}_N$ containing $N$ referred users which are the most similar ones to user $U$. The criterion for the selection is given by:
\begin{equation}
    \mathcal{R}_N={\rm argmax}_{R'\subset R, |R'|=N}\sum_{\mathbf{r}\in R'} sim(\mathbf{u},\mathbf{r})
    \label{eqn:TopN}
\end{equation}

The estimated output features are calculated by:
\begin{equation}
\widehat{\mu}_{u,p} =  \frac{\sum_{r \in \mathcal{R}_N}\mu_{r,p} sim(\boldsymbol{\mathbf{u}},\boldsymbol{\mathbf{r}})}
{\sum_{r \in \mathcal{R}_N}sim(\boldsymbol{\mathbf{u}},\boldsymbol{\mathbf{r}})}
\label{eqn:estMu}
\end{equation}

\subsection{Recommendation system-based UCB}
We assume that a rich historical dataset containing previous (including both organic and bandit) sessions on the products of several users is available. In order to perform our proposed recommendation system-assisted UCB, the historical dataset is filtered to formulate a reference set.
The users that are selected to enter the reference set should satisfy at least one of the two following conditions: (i) the total length of both their organic and bandit sessions is long enough, and (ii) they have positive interactions (i.e. click) on bandit sessions.
The first condition ensures that the record contains enough information about the latent interest of users about the products. Indeed, with a short record of either organic or bandit session, we cannot conclude anything about the interests of a user. 
The second condition enables us to keep sparse information of the positive interaction on bandit session. Due to the nature of advertisement problem, the number of positive interactions (i.e., clicking on the recommended products) are often very low. Thus keeping users' record with positive interactions on bandit session can help to enrich the knowledge of product-product and product-user correlations. 

\begin{algorithm}[htp]
\SetAlgoLined
\KwResult{Recommended Product $I_t$}
    Given Reference set $\mathcal{R}$\;
    \eIf{New user}{
        Select randomly $N$ referred users from $\mathcal{R}$\;
    }{
        Compute similarity distances of the user and all of the referred users based on \eqref{eqn:cosineSim}\;
        Get $N$ top similar user $\mathcal{R}_N$ based on \eqref{eqn:TopN}\;
    }
    Compute estimated mean reward based on \eqref{eqn:estimateCTR}\;
    Compute Upper-Confidence Bound of Products based on \eqref{eqn:ModifiedUCB} \;
    $I_t \leftarrow \arg \max_i {\tilde{\mu}_{i,t}}$
 \caption{UCB-RS}
 \label{alg:UCB-RS}
\end{algorithm}

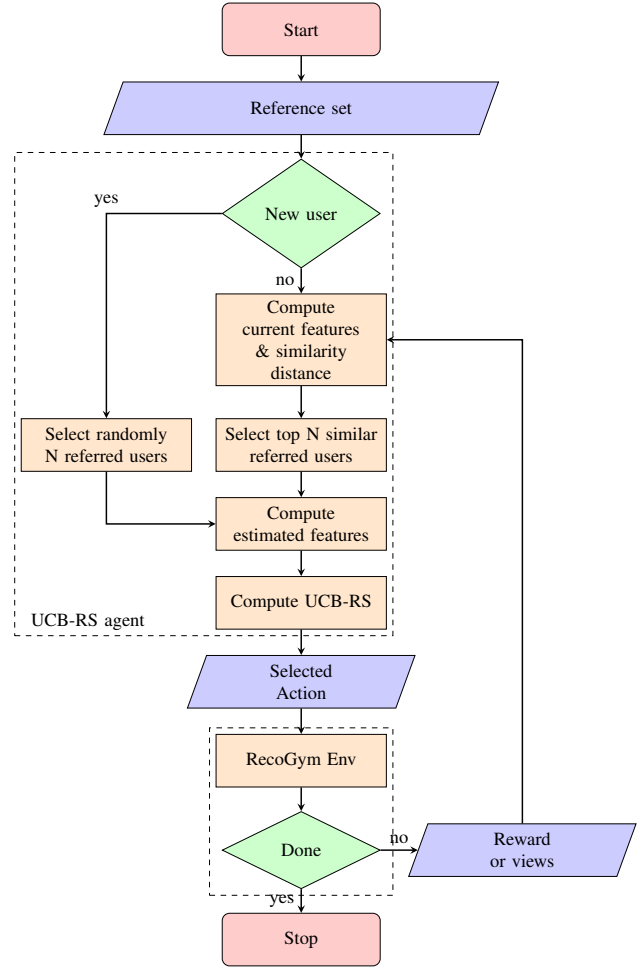
\begin{figure}[htbp]
\scalebox{.7}{
\begin{tikzpicture}[node distance=1.5cm]
\node (start) [startstop] {Start};
\node (refset) [io, below of=start]{Reference set};
\node (checknew) [decision, below of=refset,yshift=-0.5cm] {New user};
\node (pro1) [process, below of=checknew,yshift=-.9cm] {Compute current features \\ \& similarity distance};

\node (pro2b) [process, below of=pro1,yshift=-0.5cm] {Select top N similar referred users};
\node (pro2a) [process, left of=pro2b,xshift=-2.2cm] {Select randomly N referred users};
\node (pro3) [process, below of=pro2b] { Compute estimated features};
\node (pro4) [process, below of=pro3] { Compute UCB-RS };
\node (action) [io, below of=pro4] {Selected \\ Action};
\node (reco) [process, below of=action] { RecoGym Env };
\node (done) [decision, below of=reco,yshift=-0.2cm] {Done};
\node (stop) [startstop, below of=done,yshift=-0.2cm] {Stop};
\node (reward) [io, right of=done,xshift=2.7cm] {Reward \\ or views};
\node[draw,rectangle,very thin, dashed,fit=(checknew)(pro1) (pro2a)(pro2b)(pro3)(pro4),
label={[shift={(-2.2,-9.2)}]UCB-RS agent}] (rec) {} ;
\node[draw,rectangle,very thin, dashed,fit=(reco) (done)] (rec) {} ;

\draw [arrow] (start) -- (refset);
\draw [arrow] (refset) -- (checknew);
\draw [arrow] (checknew) -- node[anchor=east] {no} (pro1);
\draw [arrow] (checknew) -| node[anchor=south] {yes} (pro2a);
\draw [arrow] (pro1) -- (pro2b);
\draw [arrow] (pro2b) -- (pro3);
\draw [arrow] (pro2a) |- (pro3);
\draw [arrow] (pro3) -- (pro4);
\draw [arrow] (pro4) -- (action);
\draw [arrow] (action) -- (reco);
\draw [arrow] (reco) -- (done);
\draw [arrow] (done) -- node[anchor=south] {no} (reward);
\draw [arrow] (done) -- node[anchor=east] {yes} (stop);
\draw [arrow] (reward) |- (pro1);
\end{tikzpicture}
}
\caption{UCB-RS on Reco-gym}
\label{fig:UCB_RS}
\end{figure}

The algorithm of the proposed UCB-RS is presented in Alg.~\ref{alg:UCB-RS} and Fig.~\ref{fig:UCB_RS}. The initialization of UCB-RS algorithms requires a reference set. At the initial step, the algorithm computes the instant feature vector of the current user. It could be either the click through rate or the probability of viewing products, and represents the characteristics of the investigated user. The feature vector is then used to calculate the similarities with the referred users. Top N similar users will be selected as shown in \eqref{eqn:TopN}. We compute the estimated feature vector based on \eqref{eqn:estMu}. The output of UCB-RS is accordingly given by:
\begin{equation}
    U_{k,t} = \tilde {\mu}_{i,t} + \xi_{i,t} 
    \label{eqn:ModifiedUCB}
\end{equation}
where $\xi_{i}$ is the confidence interval and is defined by:
\begin{equation}
    \xi_{i,t} = 
        \left\{
            \begin{matrix}
                \sqrt{\alpha \log {T}}&  \textup{ if } N_i = 0 \\
                \sqrt{\frac{\alpha \log {t} }{N_i}}& \textup{ otherwise }
            \end{matrix}
        \right.
    \label{eqn:ConfidentInterval}
\end{equation}
$T$ is a predefined time interval which is much longer than the length of most of the bandit sessions.
$\tilde {\mu}_{i,t}$ is the estimated average reward of arm $i$ at time $t$, and is heuristically defined by:
\begin{equation}
     \tilde {\mu}_{i,t} = \lambda {\mu}_{i,t} + (1-\lambda)  \hat {\mu}_{i,t}
     \label{eqn:estimateCTR}
\end{equation}
where $0\leq \lambda \leq 1$, ${\mu}_{i,t}$ is the instant average reward of arm $i$ at time $t$, and $ \hat {\mu}_{i,t}$ is the estimated mean reward of arm $i$ obtained by the recommendation system in \eqref{eqn:estMu} .

It can be seen that the confidence interval given in \eqref{eqn:ConfidentInterval} is almost similar to that given in \eqref{Eqn:UCB}. The reason behind this utilization is that we use the same instant arms' average reward values for computing the referred similarities and then the estimated mean rewards.
The difference occurs only when we calculate the confidence interval of arm $i$ which has never been played yet, i.e., $N_i=0$. This case does not happen in the original UCB1 algorithm because its initial stage requires us to play  each arm once before delving into the algorithm. Note that this is the main drawback of UCB1 when adopted to a very large arm space. In our proposed UCB-RS, thanks to the recommendation system, the initial stage is ignored and hence, to calculate the confidence interval of an unexplored arm, we assume that it had been played only one time for a very long duration $T$.

\section{Numerical results} \label{sec:Result}
In order to investigate the effectiveness of the proposed method, we build up our agents and other baseline comparing agents. The agents are then tested with RecoGym environment. We configure the environment based on the parameters given in Table \ref{Tbl:Params}.

\begin{table}[htb]
\begin{tabular}{ll}
\hline
\multicolumn{1}{c}{\textbf{Parameter}}                        & \textbf{Value} \\ \hline
Size of users latent representation of interests                & K=10           \\ 
Std. deviation of users latent representation of interests  & $\sigma_\Omega = 1$              \\ 
Std. deviation of items latent representation of popularity & $\sigma_\mu = 3\to30$           \\ 
Transition Prob. from bandit session to organic session         & 0.05,          \\ 
Transition Prob. from organic session to bandit session         & 0.25,          \\ 
Transition Prob. of leaving bandit session                      & 0.01,          \\ 
Transition Prob. of leaving organic session                     & 0.01,
        \\ 
Number of random users for selecting reference set           & 2000,
        \\ 
Number of tested users                     & 100,
        \\ \hline
\end{tabular}
\caption{Simulation Parameters}
\label{Tbl:Params}
\end{table}

To compute the estimated mean reward $ \hat {\mu}_{i,t}$, there are a lot of possible estimating combinations. We first consider the two following cases  to observe the separative influence of RS on the final performance:
\begin{itemize}
    \item UCB-RS I - Features: instant ctr $\chi_{i,t}$ of user at arm $i$-  $ \hat {\mu}_{i,t}$: Combined ctr of top-N referred users.
    \item UCB-RS II - Features: instant probability of viewing item $i$ in organic sessions $\nu_{i,t}$ of user - $ \hat {\mu}_{i,t}$: Combined ctr of top-N referred users.
\end{itemize}

Secondly, we introduce UCB-RS III a heuristic proposal utilizing both instant \textit{ctr} and instant probability of viewing item $i$ in organic sessions of user as separate features. The estimated average reward given in \eqref{eqn:estMu} is computed by:
$$ \hat {\mu}_{i,t} = \lambda(\chi_{i,t} + \theta \nu_{i,t} ) + (1-\lambda) (\hat{\chi}_{i,t} + \theta \hat{\nu}_{i,t} )$$
where $\theta$ is a scale factor.

\subsubsection{Comparison among MAB algorithms}
\begin{table*}[hbt]
    \centering
    \begin{tabular}{lccccccc}
    \toprule
     & $\epsilon$-Greedy &       EXP3 &      EXP3S &       UCB1 & UCB-RS I & UCB-RS II & UCB-RS III  \\
    \midrule
     $\sigma_\mu$= 3 & 1.1612 &   1.1262 &   1.2098 &  1.3437 &   1.3475 &\textit{1.5961} & \textbf{1.7603}\\
     $\sigma_\mu$= 5 & 1.4377 &   1.1213 &   1.4403 &  1.4202 &   \textit{1.5576}
 & 1.3702 & \textbf{1.8032}
\\
     $\sigma_\mu$=10 &         1.2881 &  1.2136 &  1.3323 &  1.3941 &      1.4256 & \textit{2.0921} &  \textbf{ 2.2051}
\\
     $\sigma_\mu$=15 &         1.3304 &  1.1128 &  1.2409 &  1.3886  &      1.4619
&\textit{ 1.7795} &  \textbf{ 2.3964}
\\
     $\sigma_\mu$=20 &         1.7147 &  1.2586 &  1.4129 &  1.3853 &       1.7266
 & \textit{2.1865} &   \textbf{2.3522}
 \\
     $\sigma_\mu$=25 &          1.3146 &  1.1607 &  1.5691 &  1.4734 &      1.6539
 &\textit{2.0928}  & \textbf{2.5476}
 \\
     $\sigma_\mu$=30 &         1.3021 &  1.1467 &  1.6719 &  1.4711 &    1.8919
 & \textit{2.2041} &   \textbf{2.6706}
\\
    \bottomrule\end{tabular}
    \caption{Average testing click through rate [\%] with 50 products case}
    \label{tab:ctrAll}
\end{table*}
\begin{figure*}[htbp]
\begin{subfigure}{.33\textwidth}
  \centering
  \includegraphics[width=\textwidth]{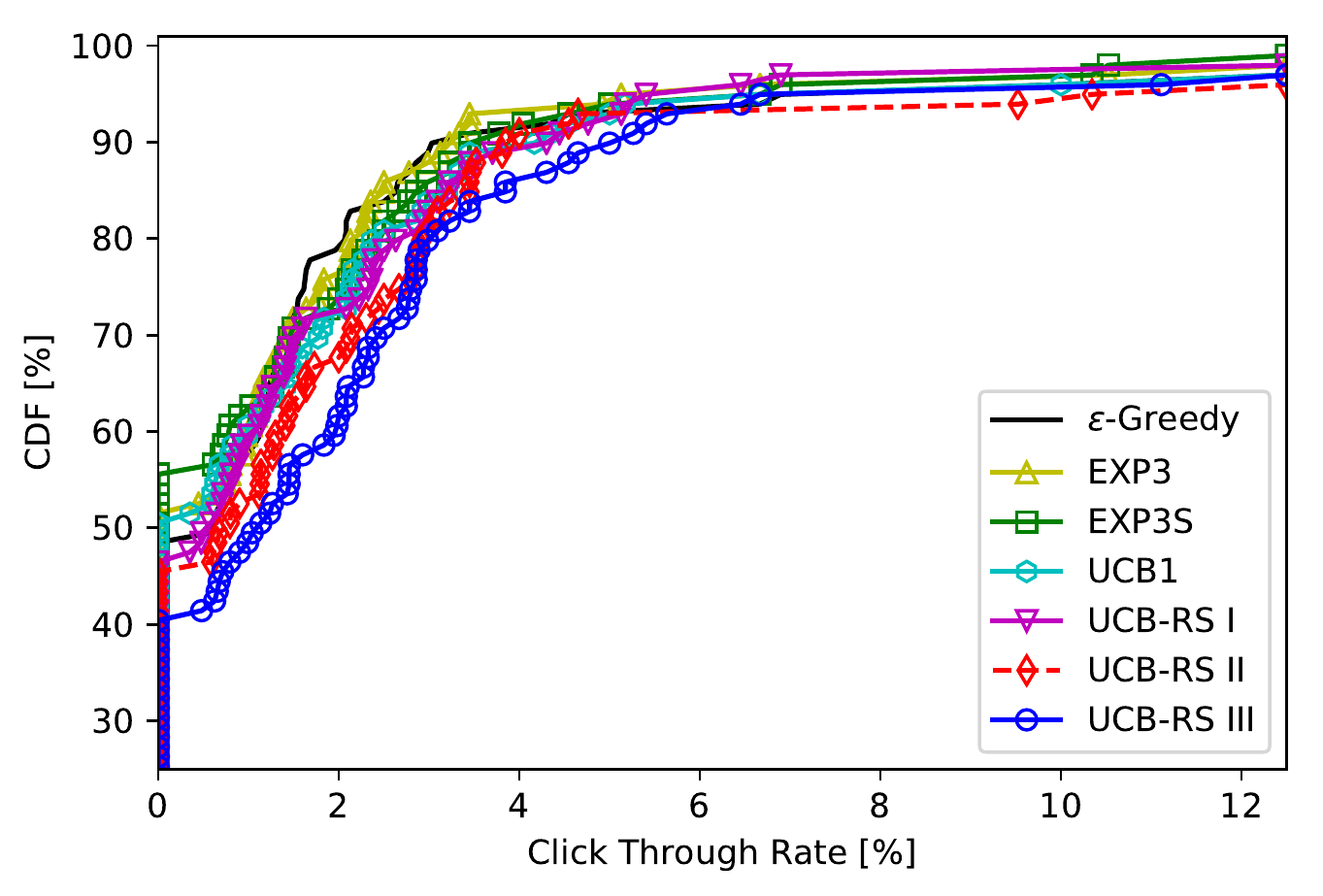} 
        \caption{ $\sigma_{\mu} = 3$}
        \label{fig:CDF_Sigma_3}
\end{subfigure}%
\begin{subfigure}{.33\textwidth}
  \centering
  \includegraphics[width=\textwidth]{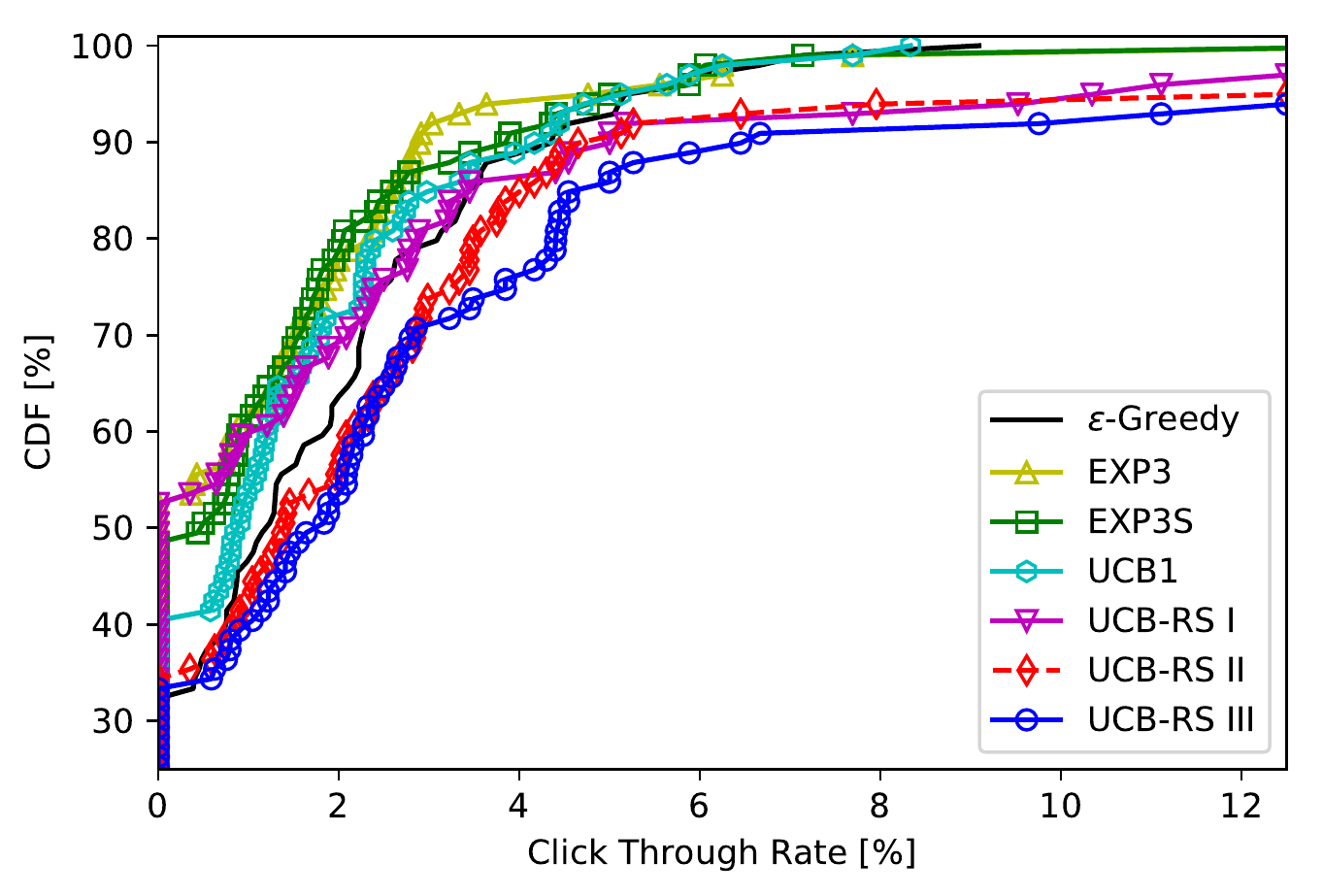} 
        \caption{ $\sigma_{\mu} = 10$}
        \label{fig:CDF_Sigma_10}
\end{subfigure}
\begin{subfigure}{.33\textwidth}
  \centering
  \includegraphics[width=\textwidth]{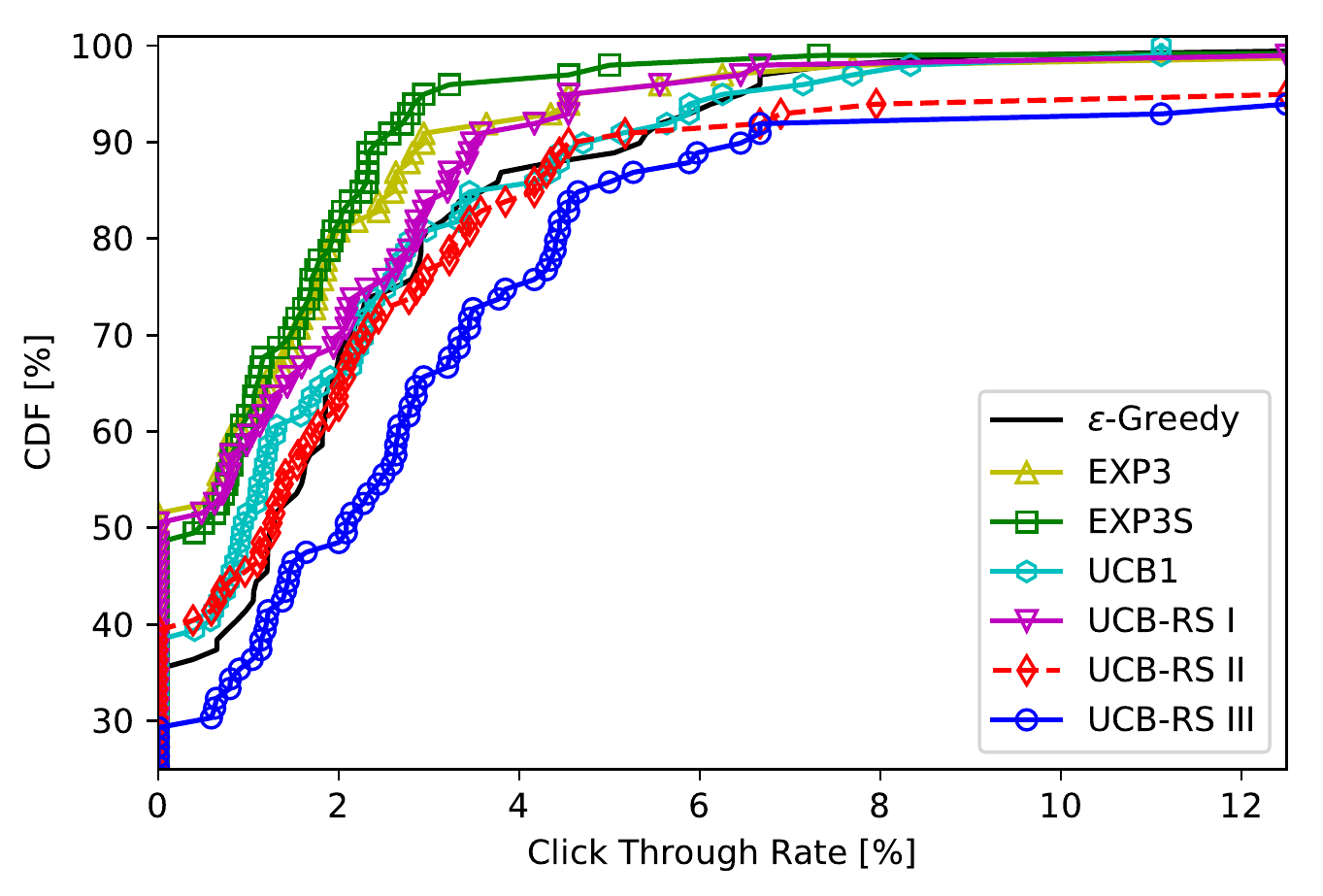} 
        \caption{ $\sigma_{\mu} = 15$}
        \label{fig:CDF_Sigma_15}
\end{subfigure}%

\begin{subfigure}{.33\textwidth}
  \centering
  \includegraphics[width=\textwidth]{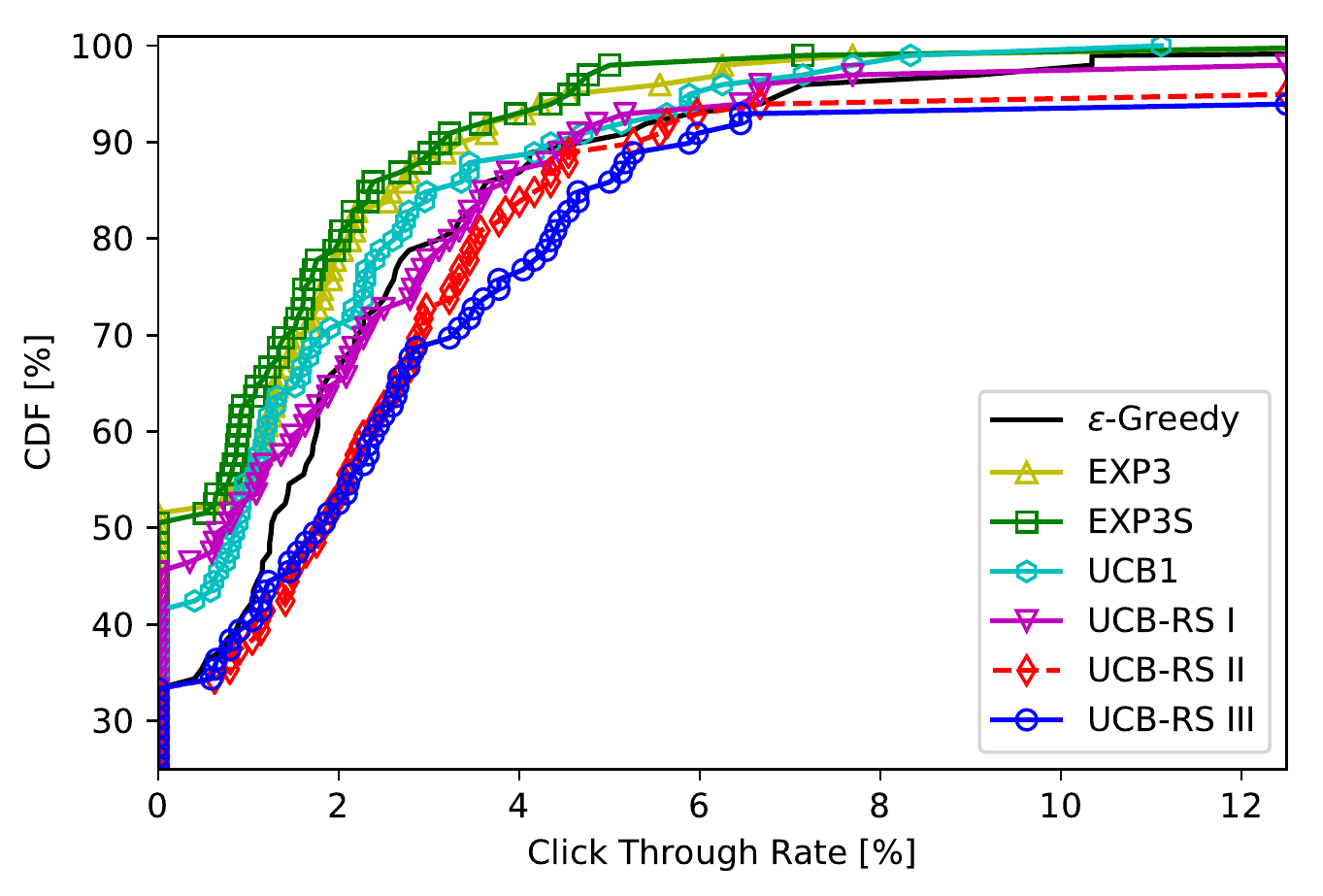} 
        \caption{ $\sigma_{\mu} = 20$}
        \label{fig:CDF_Sigma_20}
\end{subfigure}
\begin{subfigure}{.33\textwidth}
  \centering
  \includegraphics[width=\textwidth]{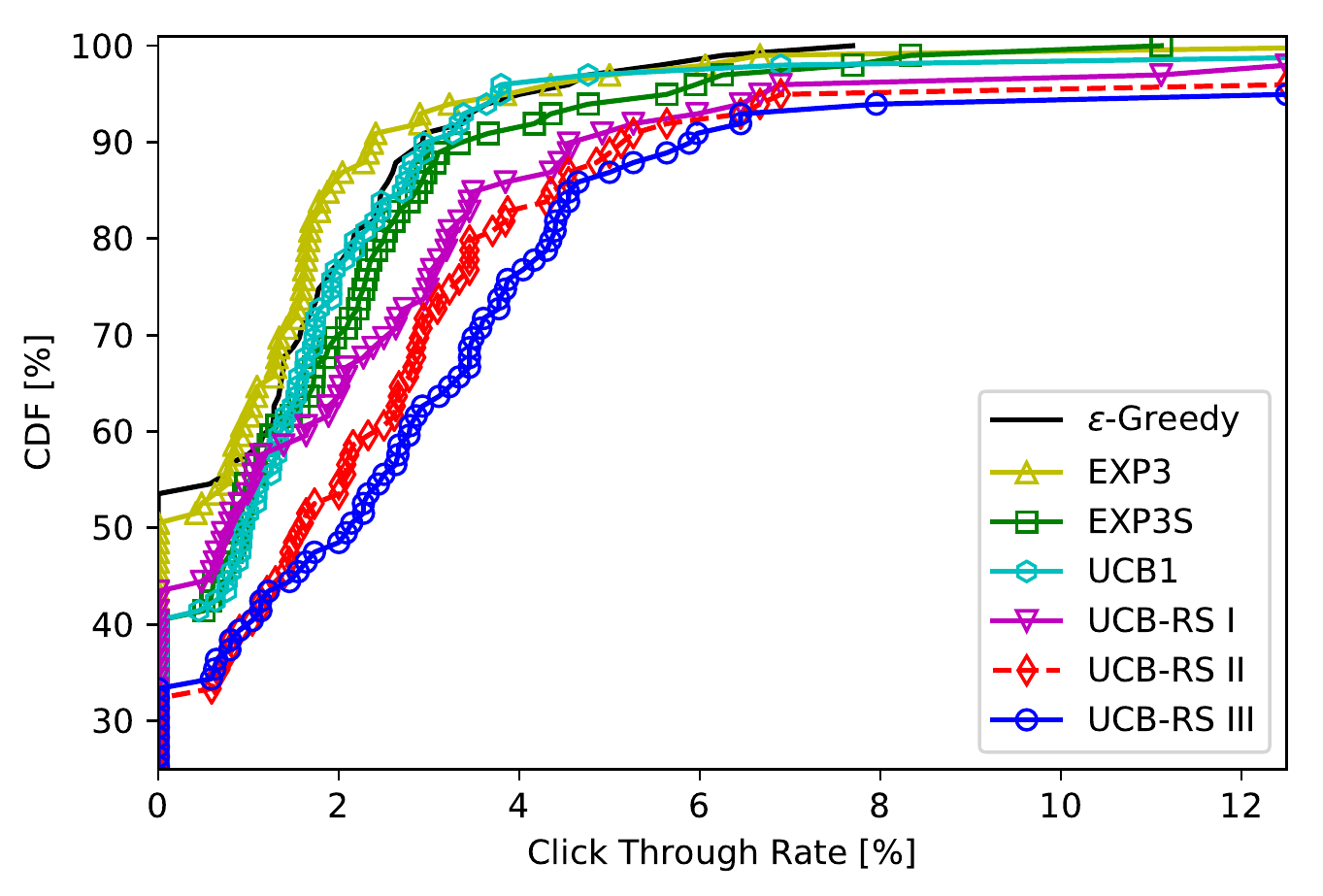} 
        \caption{$\sigma_{\mu} = 25$}
        \label{fig:CDF_Sigma_25}
\end{subfigure}
\begin{subfigure}{.33\textwidth}
  \centering
  \includegraphics[width=\textwidth]{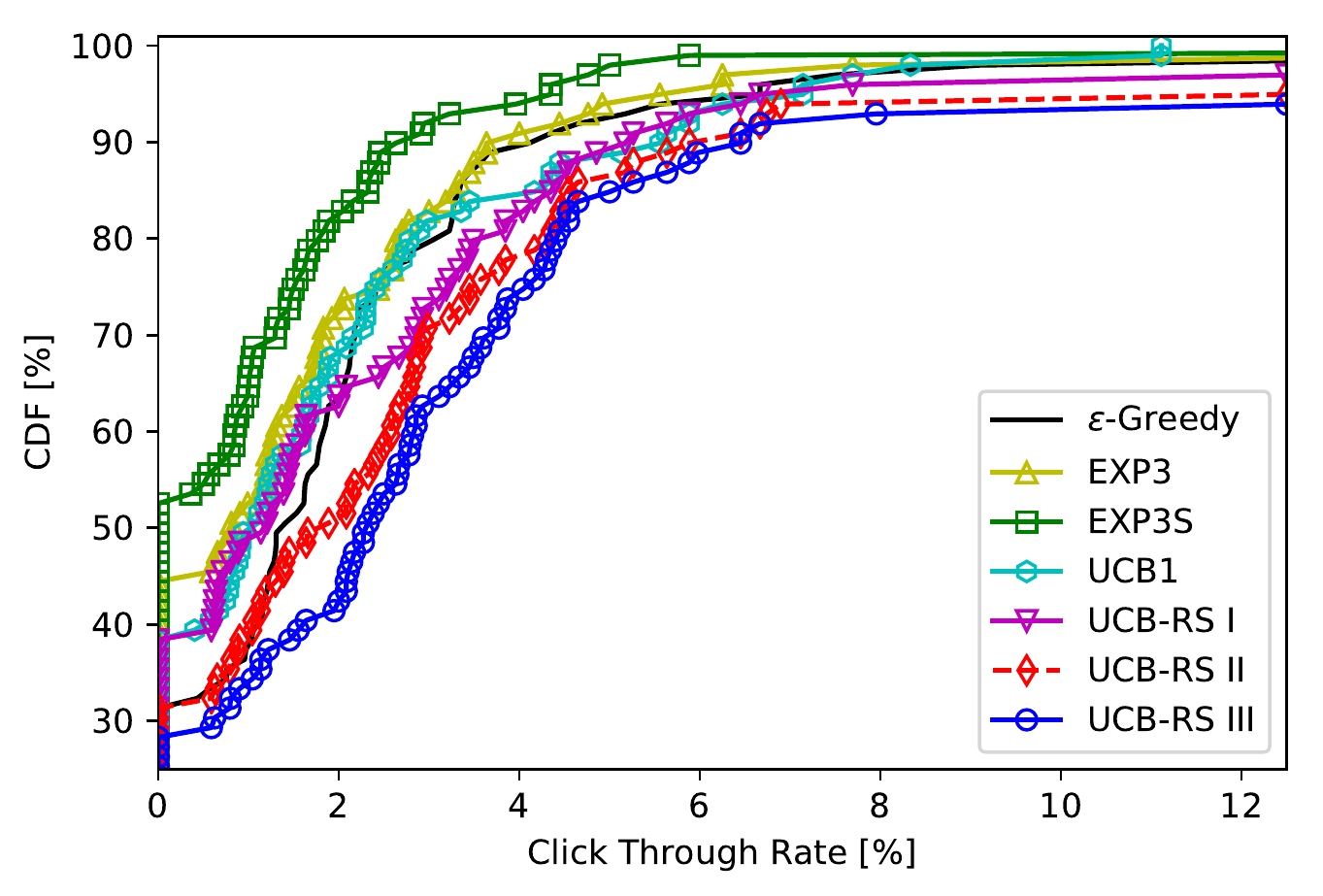} 
        \caption{ $\sigma_{\mu} = 30$}
        \label{fig:CDF_Sigma_30}
\end{subfigure}

\caption{CDF with different values of  $\sigma_{\mu}$ with 50 products}
\label{fig:CDF}
\end{figure*}

First, we compare the proposed methods with the conventional ones including $\epsilon$-Greedy, EXP3, EXP3S, UCB1 under the same environment condition. The results are shown in Table \ref{tab:ctrAll}.
We can see that the proposed UCB-RS III, using both the instant bandit click through rate and the organic probability of product view as the features for the recommendation system, outperforms all other algorithms. When using separately each for the two features, i.e. UCB-RS I and II, our proposed method also performs best in most of the cases. Comparing between UCB-RS I and UCB-RS II, it can be seen that the second method, which uses the probability of product view in organic sessions as the features of the recommendation system, works better than the first one. We can explain this by the obtained diversity of information for making the decision for product recommendation. In UCB-RS I, despite the correlation information about the product rewards taken from other referred users through the mechanism of recommendation, the second term of the sum given in \eqref{eqn:estimateCTR} derives from the first one. In other words, there is a strong correlation between the two terms of \eqref{eqn:estimateCTR} in this case because they are all the information available in bandit sessions. Contrarily, for UCB-RS II, the feature for the recommendation frame work is now the probability of products view in organic sessions. This makes the estimated mean rewards given in  \eqref{eqn:estimateCTR} contain more diverse information, and hence makes the UCB-RS agent performs better. 
In summary, it can be seen the proposed UCB-RS III gain from 20\% to 40\% of click through rate compared to UCB1 and EXP3S. Noticed that EXP3S and UCB1 perform better than other the conventional RL schemes. 

The results given in Table \ref{tab:ctrAll} also show us that the performance of our proposed UCB-RS depends on the value of $\sigma_\mu$, which is not the case of other methods. Recall that $\sigma_\mu$ is the standard deviation of the latent popularity of products. This means that the larger $\sigma_\mu$ is, the more the product set contains popular items. Therefore, it can be seen that the recommendation framework in our proposed method works more effectively with the increasing number of popular products.



Secondly, we present the results given in Table \ref{tab:ctrAll} in a detailed view in Fig. \ref{fig:CDF}. The curves represent the cumulative distribution function (CDF) of the click through rates of the tested users. For example, in Fig. \ref{fig:CDF_Sigma_15}, it can be seen that for a click through rate of $2\%$, our proposed method UCB-RS III has $>50\%$ of tested users, whereas the other schemes have $<40\%$ tested users obtaining more than this value. For all $\sigma_\mu$ adopted, the curves of UCB-RS III tops all the others. This remark is similar to what we have seen from Table \ref{tab:ctrAll}.

\subsubsection{The impact of $\lambda$ and top-$N$}
Thirdly, we investigate the influence of $\lambda$ given in \eqref{eqn:estimateCTR} and the number of the most similar referred users top-$N$. The proposed UCB-RS III agent is tested with the case of 50 products and $\sigma_\mu$ = 10.

 \begin{figure}[htb]
\centering
  \includegraphics[width=0.475\textwidth]{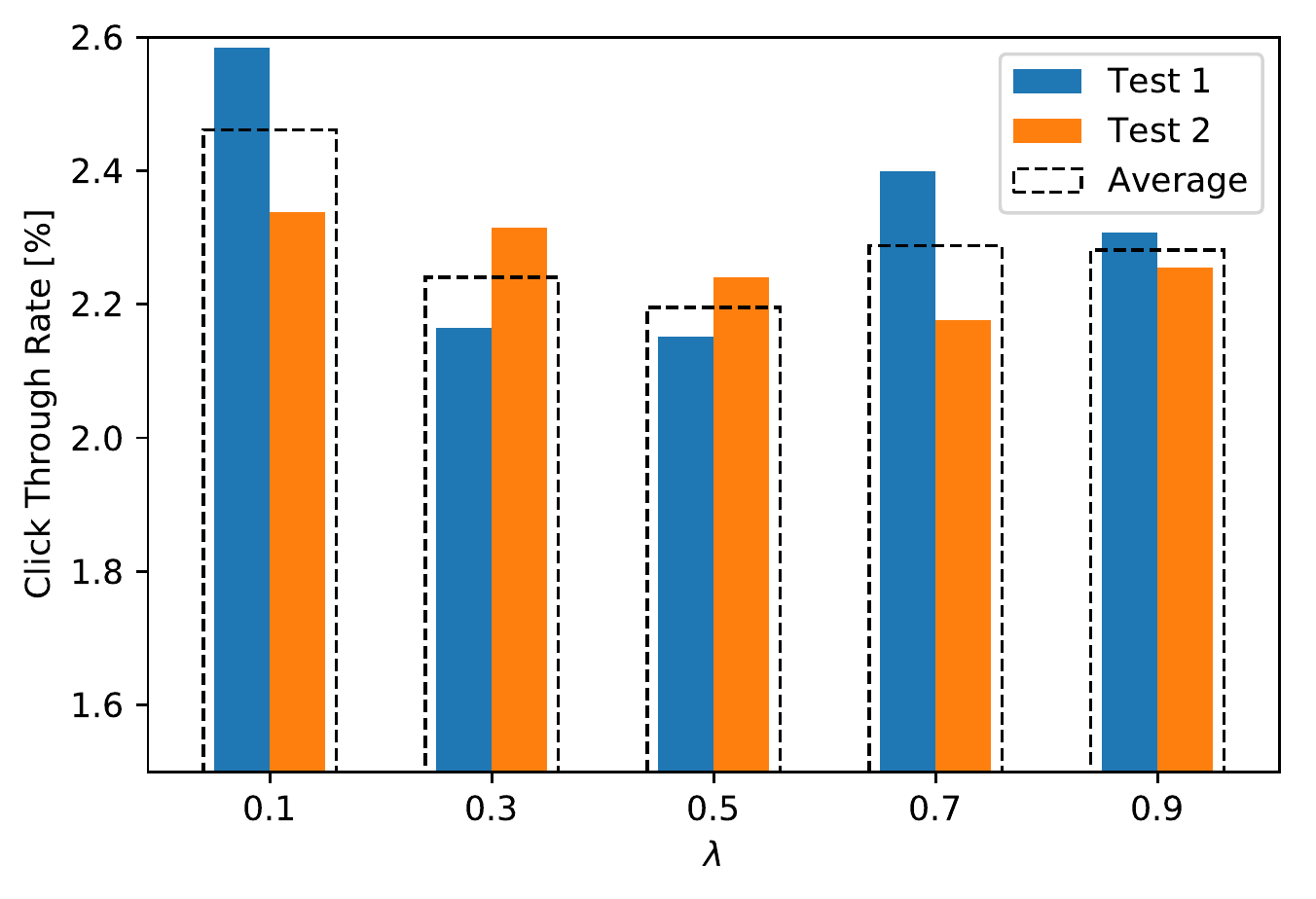}
        \caption{Click Through Rate vs. $\lambda$ when top-N = 15, $\sigma_\mu$ = 10 with 50 products}
        \label{fig:lambda}
 \end{figure}
 
 Fig. \ref{fig:lambda} shows the achieved click though rate of the agent for various values of $\lambda$ (two running times for each value of $\lambda$) when top-$N$=15. We can observe that the output performance depends on the value of $\lambda$. However, the difference performance gap is small. 
 
 \begin{figure} [htb]
\centering
  \includegraphics[width=0.475\textwidth]{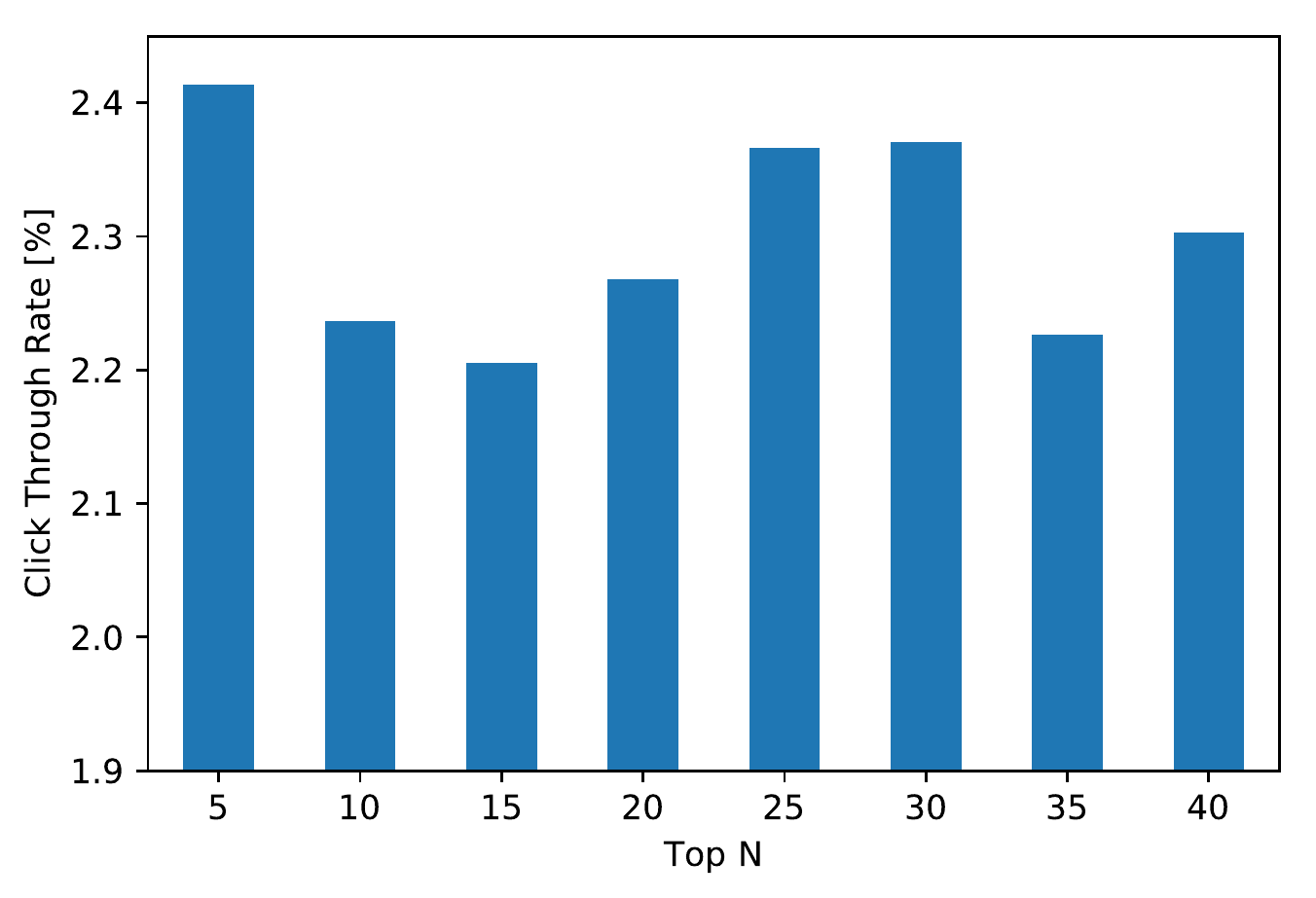}
        \caption{Click Through Rate vs. top-N when $\lambda$ = 0.5, $\sigma_\mu$ = 10 with 50 products}
        \label{fig:TopN}
 \end{figure}
 
 Fig.  \ref{fig:TopN} shows the achieved click though rate of the agent for different values of top-$N$ when $\lambda = 10$. It can be seen that varying the value of top-$N$ affects the output click through rate of the agent. However, even in the smallest achieved click through rate case, i.e. top-$N$=15, the proposed UCB-RS III outperforms the other algorithms as shown in the third row of Table \ref{tab:ctrAll}.

\subsubsection{The impact of the number of products}
\begin{figure}[htb]
    \begin{subfigure}{0.45\textwidth}
        \centering
        \includegraphics[width=\textwidth]{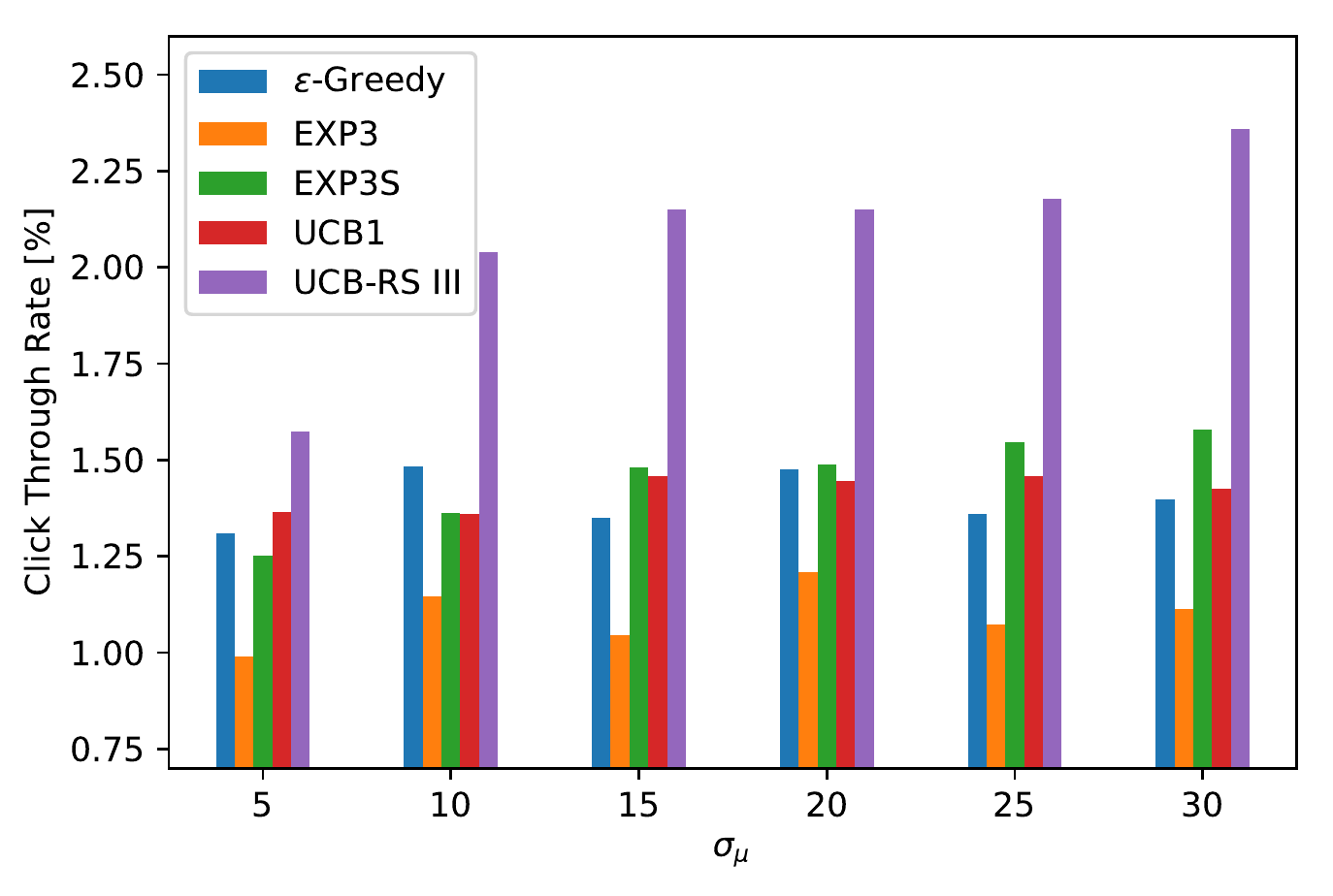}
        \caption{}
        \label{fig:BarCtrProd100}
    \end{subfigure}
    \begin{subfigure}{0.45\textwidth}
        \centering
        \includegraphics[width=\textwidth]{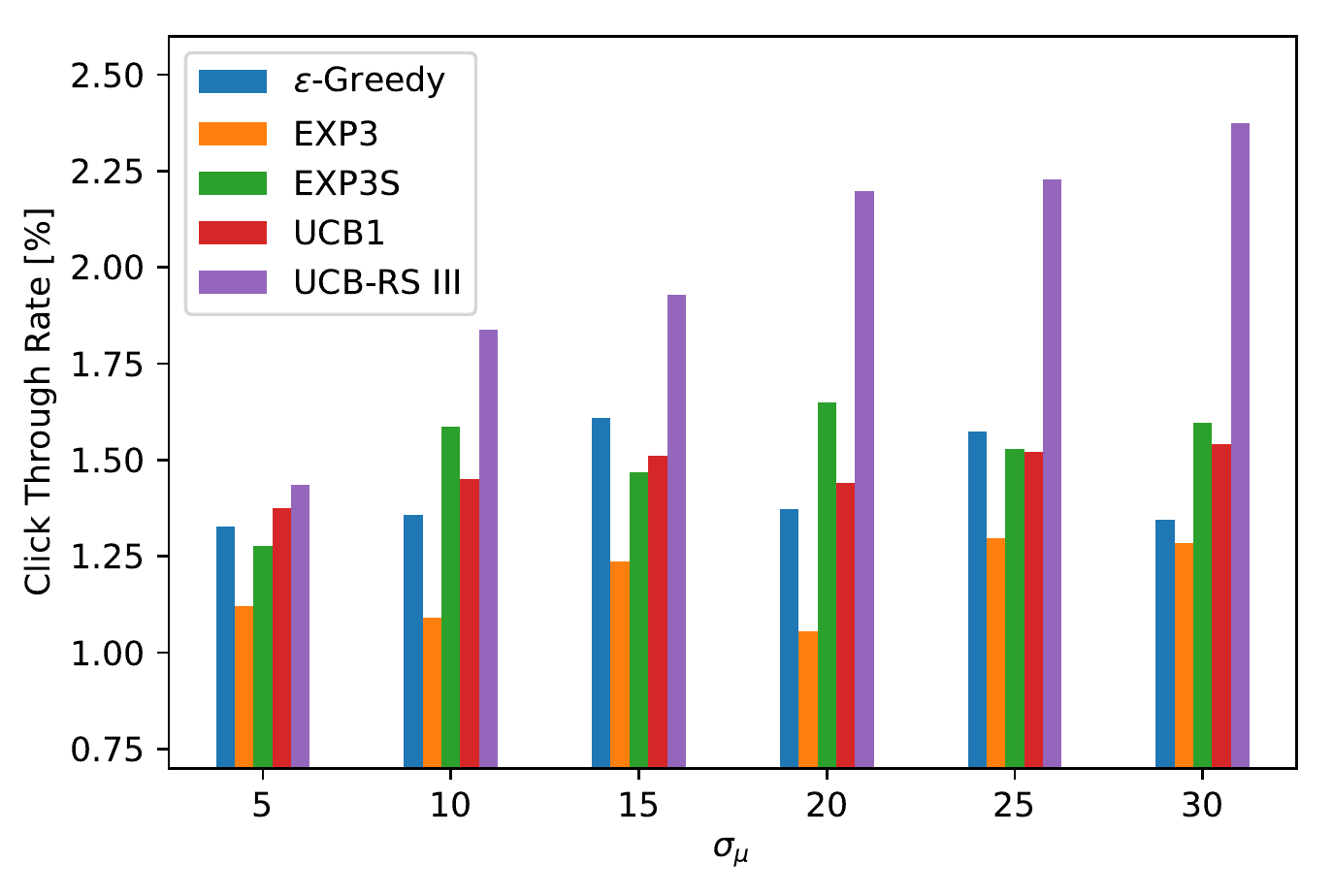}
        \caption{}
        \label{fig:BarCtrProd200}
    \end{subfigure}
    \caption{Click Through Rate vs. methods when top-N = 15, $\sigma_\mu$ = 15, $\lambda=0.5$ with (a) 100 and (b) 200 products}
    \label{fig:BarCtrProd}
\end{figure}

Finally, we investigate the impact of the number of the products on the achieved click through rate of the proposed method. The results are shown in Fig. \ref{fig:BarCtrProd}.

 Comparing with the click through rate of UCB-RS III obtained in the 100 products case shown in Fig. \ref{fig:BarCtrProd100}, the click through rate of UCB-RS III in the 200 products case shown in Fig. \ref{fig:BarCtrProd200} has a small trend of decrease. However, compared with other schemes with any adopted  $\sigma_\mu$ and both 100 and 200 products, our proposed UCB-RS III obtains the highest click through rate similar to the results obtained with 50 products shown in Table \ref{tab:ctrAll}.

 \section{Conclusion} \label{sec:Conclusion}
In this paper, we have devised the UCB-RS method, which uses recommendation system for enhancing the upper-confidence bound algorithm. The proposed method is targeted to the product recommendation problem in online advertising. 
Our goal is to improve product recommendation by dealing with non-stationary and large state spaces multi-armed bandit problems. By testing with RecoGym, an OpenAI Gym-based reinforcement learning environment for the product  recommendation in online advertising, we have shown that the proposed method can outperform widespread reinforcement based algorithms such as $\epsilon$-Greedy, UCB1, EXP3 and EXP3S. The results show that the proposed method gains from 20\% to 40\% of click through rate compared to the conventional UCB1 algorithm. Though achieving potential results, the analysis of the proposed method in terms of the theoretical performance, the complexity and the scale-ability needs to be considered in the future. Other possible features for the RS should also be investigated more.


\bibliographystyle{IEEEtran}
\bibliography{RSUCB}

\end{document}